\documentstyle[prb,aps,epsfig,twocolumn]{revtex}
\begin{document}
\draft
\flushbottom
\twocolumn[
\hsize\textwidth\columnwidth\hsize\csname @twocolumnfalse\endcsname
\title{Level curvatures, spectral statistics and scaling  
for interacting particles.} 
\author{Eric Akkermans\cite{ea} and Jean-Louis Pichard\cite{jlp} 
\\Institute for Theoretical Physics, University of California, 
Santa Barbara, Ca 93106-4030, USA}
\maketitle
\tightenlines
\widetext
\advance\leftskip by 57pt
\advance\rightskip by 57pt
\begin{abstract}
  The mobility of two interacting particles in a random potential 
is studied, using the sensitivity of their levels to a change 
of boundary conditions. The delocalization in Hilbert space induced 
by the interaction of the two particle Fock states is shown to decrease the 
mobility in metals and to increase it in insulators. In contrast to 
the single particle case, the spectral rigidity is not directly 
related to the level curvature. Therefore, another curvature of 
topological origin is introduced, which defines the energy scale 
below which the spectrum has the universal Wigner-Dyson rigidity. 
\end{abstract}


]
\narrowtext
\tightenlines

\setcounter{equation}{0}
\def\real{{\rm I\kern-.2em R}}
\def\complex{\kern.1em{\raise.47ex\hbox{
	    $\scriptscriptstyle |$}}\kern-.40em{\rm C}}
\def\integer{{\rm Z\kern-.32em Z}}

%           ================ INTRODUCTION ================
%

 The interplay between disorder and interactions is a central problem in 
quantum transport. Recently, it was proposed \cite{Shepelyansky} 
that interactions could favor delocalization in disordered insulators 
of large (one particle) localization length $L_1$. This was 
first understood for the simple case of two interacting particles (TIP) 
in a random potential, where a fraction of the TIP-states do have 
~\cite{Shepelyansky} a localization length $L_2$ larger than $L_1$. This 
result is surprising since in disordered metals, repulsive 
interactions are expected to reduce
 the conductivity~\cite{Altshuler} and in disordered 
insulators, the singularity of the density of states which leads eventually 
to a gap, also reduces the quantum transport. This delocalization effect 
can be understood by considering the problem in the Fock space, 
where the many-body states are delocalized by the interaction. 
This is a very general idea that we use here in the restricted framework of 
the two-body problem. By Fock states, we simply mean the $2 \times 2 $ 
Slater determinants built out of the one particle states, assuming  
spinless fermions. These are eigenstates in the absence of 
interaction, and form a basis of the TIP-Hilbert space. 
This terminology of Fock states might not be the correct one 
for this two body problem, but we use it both for brievety and because 
it will become appropriate for the general N-body 
problem ~\cite{agkl}. We consider bare particles, 
but the extension to dressed quasi-particles created from the Fermi 
vacuum is straightforward ~\cite{Imry}. In the presence of interactions, 
characterized by the parameter $U$, the Fock states are 
broadened~\cite{js,wp}, and the TIP eigenstates have a finite 
projection over (typically) $g_2 \equiv \Gamma / \Delta_2^e$ nearby in 
energy Fock states. Here, $\Gamma$ is the 
broadening, characterizing the local density 
of states in the Fock basis (assuming a Breit Wigner form) and $\Delta_2^e$ the 
spacing between the Fock states effectively coupled by $U$. This 
delocalization in the Fock basis characterizes both metals and insulators. 
But the delocalization in the real space occurs only for insulators, 
since the Fock states are themselves localized in real space. 
 
   In order to understand this delocalization,
 a scaling approach of the TIP-problem was  
proposed ~\cite{Imry} as a generalization of the Thouless block 
scaling picture. For system sizes $L_1<L<L_2$,
a certain conductance
${g_2}(L)= {\Gamma / {\Delta_2^e}}$ was
defined and assumed to obey the usual Ohm's law
(i.e. the one particle scaling
theory of localization) in order to recover the delocalization by
interactions. In that work, as in an effective $\sigma$ model 
formulation  developed~\cite{fmgp} later, $L_1$ was implicitely 
assumed to be the smallest resolved scale.
 Therefore, the meaning of $g_2$ is unclear 
for $L\ll {L_1}$ i.e. in the metallic regime.

Independently, an analysis~\cite{wp} of the statistical properties of the 
TIP-spectrum in the metallic regime has shown that 
the universal Wigner Dyson rigidity occurs for ${g_2}= 
{\Gamma \over{\Delta_2^e}}$ consecutive levels, provided $g_2 \geq 1$. 
In the absence of interaction, the spectrum is essentially of the Poisson 
type for energies $ {\Delta_2} \leq E \leq {\Delta_1}$ where ${\Delta_1}$
is the one particle mean level spacing. For the one particle spectrum, 
one knows that Wigner-Dyson rigidity occurs for $g_1$ consecutive levels, 
where $g_1$ is the dimensionless conductance of the non interacting electron 
gas. For the TIP-case, $g_2$ plays the role of $g_1$, as far as the 
spectral fluctuations are concerned. We will show in this letter that, 
though the TIP-spectrum becomes more rigid for stronger interactions, 
the transport in real space is reduced when $L<L_1$.

 To that purpose, we describe the
transport in terms of the sensitivity of the
levels to a change of boundary conditions
 ~\cite{Thouless}.
This is equivalent to closing the system as a ring of length L
pierced by a dimensionless Aharonovh  -Bohm magnetic flux
 $\phi = {\Phi / {\Phi_0}}$
where ${\Phi_0}= {{hc}/ e}$. We
define the two-particle level curvature ${C_2}(E)$ at energy $E$ as the
zero flux curvature of the integrated  two-particle density of states
$N(E,\phi)$:
\begin {equation}
{C_2}(E)={\partial^2 \over \partial\phi^2} N(E,\phi){\big\vert_{\phi=0}}
\end{equation}

An equivalent expression was shown ~\cite{am} to describe the d.c residual
dimensionless
conductance $g_1$ of the non interacting electron gas under very general
conditions set by the random  matrix theory. It is therefore
natural to try to extend this description to the case of interacting particles.

The main results of this letter are as follows. 
Assuming a Breit-Wigner form for the local density of states
~\cite{js,wp}, we obtain for the two particle curvature $C_2$ the expression
\begin{equation}
C_2(E)= {{C_2}^{(0)}}(E) 
- {{g_2} \over {\Delta_1}} I({E \over \Gamma},{B \over \Gamma})
\end{equation}
where ${{C_2}^{(0)}}(E)$ is the value of the curvature in the absence 
of interactions, $B$ the kinetic energy scale (bandwidth)
and $I({E \over \Gamma},{B \over \Gamma}) \approx  
{{{g_1}{\Delta_1}} \over B}$ for ${g_1} \gg 1$ and of order $-\Delta_1$ for 
${g_1} \ll 1$. Since $C_2(E)$ fluctuates from sample to sample,
 let us make more precise what is the meaning of Eq.(2).  For the one 
particle case, the curvature $C_1(E)$ is 
characterized, in the metallic regime, by a very
broad distribution (generalized Lorentzian) 
~\cite{delande}. Its mean value over the statistical
ensemble is zero at the center of the
band ($E=0$) so that $g_1$ is typically given by the width of the 
distribution, defined for
instance by the well behaved quantity $<|C_1(E)|>$.
 $C_2(E)$ should be understood in a similar way i.e. as
a typical value characterizing the width of a
distribution, and not as a mean value.
Preliminary numerical results~\cite{w} display a similar behavior for the 
TIP spectrum for a large enough $U$.

In the metallic regime, the typical values of ${{C_2}^{(0)}}(E)$ and $I$ 
 are given  respectively by
$ g_1 {\Delta_1 \over {{\Delta_2}^e}}$ and
 $ g_1{{\Delta_1}\over B}$ so that the total curvature is a decreasing
function
of the interaction through the term $\propto {g_2}{g_1}$ as expected in a
good metal. In the insulating regime $({g_1} \ll 1)$,
${{C_2}^{(0)}}(E)$ is zero up to exponentially small terms 
while the typical value of $I$ is proportional to $-{\Delta_1}$ so that
 $C_2$
increases with ${g_2}$ in agreement with the delocalization effect
described above ~\cite{Shepelyansky}.

 Although the physical meaning of $C_2$ is clear, this
quantity,  unlike the case of non interacting electrons in a random
potential, can hardly be considered to be relevant for a
 scaling theory of the interacting system. From the random matrix theory
point of view, the spectrum of the interacting system displays a very
unusual behaviour. Usually, the curvatures generated by varying some
system parameter also set the energy scale below which the Wigner Dyson
rigidity of the spectrum is observed. This is a central point in the
scaling
theory of localization. The fact that $g_2$ increases with the interaction 
in the metallic regime discards $C_2$ as a
candidate for scaling. In other words, we are looking for another
"curvature" which would be proportional to $g_2$ (i.e zero for the non
interacting system). 

 Without interaction $(U=0)$, the TIP-Hamiltonian is separable into two 
identical one particle Hamiltonians, i.e. $H= {h_1}+{h_2}$ with
${h_i}={{{p_i}^2}\over{2M}}
+ V(x_i)$ and where $V(x)$ is the random potential.
In other words, the two 
particles ring can be thought of as two independent one particle rings.
 For this argument to be valid, we assume
 either discernable particles, or indiscernable particles 
with additionnal quantum numbers. This leads us to 
associate distinct Aharonov-Bohm fluxes $\phi_1$ and $\phi_2$ 
to each ring. The separabilty of the TIP-Hamiltonian
is broken by the interaction 
$U({x_1}-{x_2})$. A variation of $\phi_1$ will then induce a current
in the second ring characterized by a mutual inductance.
One can extend 
this point of view to the case where a current is driven due
to the interactions between two systems, each of them characterized
by a well defined gauge field (two SNS junctions for instance).
Similarly, this two particle 
problem in $1\it d$ can be thought of
 as a single particle problem in $2\it d$, where by 
closing the $2\it d$ system on itself as a torus, 
one can again introduce $\it two$ Aharonov-Bohm fluxes $\phi_1$ and 
$\phi_2$. Without interaction, the $2 \it d$ 
character of this equivalent 
one particle model is misleading since the Hamiltonian is separable. 
But with interactions, we have a genuine
 $2 \it d$ system and a topology similar to those considered in 
the description of the integer quantum Hall effect ~\cite{stone}. 
 We thus define the spectral two-form: 
\begin{equation}
C = Im \sum_A <{{\partial A}\over {\partial {\phi_1}} } 
\big\vert {{\partial A} \over {\partial {\phi_2}}} >
\end{equation}
where $|A>$ are TIP-eigenstates of energies $E_A$. 
Using the structure of the TIP-states in the Fock basis, 
we obtain:
\begin{equation}
C(\phi_1,\phi_2) = {g_2} V(\phi_1, \phi_2) 
\end{equation}
where $V(\phi_1, \phi_2)$ is the Berry connexion whose flux 
through a closed surface in the flux space is a geometric phase 
\cite{shapere}. This relation constitutes the second result of 
this work. The remainder of this letter is devoted to establish 
and discuss them.

%The model

 We consider two particles of mass $M$ interacting through a short 
range attractive or repulsive potential $U (x_1 - x_2)$ and submitted 
to a random potential $V(x)$. Defining, 
\begin{equation} 
{\cal L}_2 (E)  = {M \over 2}\sum_{AA'}\delta (E-E_A)
{{{\big\vert <A\big\vert P\big\vert A'>\big\vert}^2}\over{E_A - E_A'}} 
\end{equation}
where ${ P}=p_1+p_2$ is the total momentum of the two
particles,
the curvature $C_2(E)$ defined by Eq.(1) rewrites:
\begin{equation}
C_2(E)= B {\big[}\sum_{A}\delta (E-E_A)
+ {\cal L}_2 (E) {\big ]}
\end{equation}
 We define the one particle states by 
$(\big\vert\alpha>,{\epsilon_\alpha})$ and the Fock states by  
$(\big\vert\alpha\beta>, \epsilon_{\alpha\beta} 
\equiv {\epsilon_\alpha}+{\epsilon_\beta})$.
To calculate ${C_2}(E)$ for the 
interacting and disordered case, we expand the TIP-states $\big\vert A>$ 
on the Fock states $\big\vert\alpha\beta>$ :
\begin{equation}
\big\vert A> = \sum_{\alpha\beta} {C_{\alpha\beta}^A}\big\vert\alpha\beta> 
\end{equation}
so that ${\cal L}_2 (E)$ contains a product of four complex amplitudes 
${C_{\alpha\beta}^A}$. To proceed further, we assume as is usual for
 disordered metals, that only those trajectories which do 
correspond to time reversed amplitudes contribute to the sum in 
${\cal L}_2 (E)$. Therefore, we keep 
only terms of the form ${{\big\vert{{C_{\alpha\beta}}^{A}}\big\vert}^2}
{{\big\vert{{C_{\alpha'\beta'}}^{A'}}\big\vert}^2}$,
i.e those obtained for $\alpha = \gamma, \beta = \epsilon $
 and $\alpha' = \gamma', \beta' = \epsilon'$. Combinations
 of the type
$\delta_{\beta\beta'}<\alpha'\big\vert{p_1}\big\vert\alpha>
 \delta_{\alpha\alpha'}<\beta'\big\vert{p_2}\big\vert\beta>$
cancel since  $<\alpha\big\vert{p_1}\big\vert\alpha> = 0$
 for zero magnetic
flux. Then, ${\cal L}_2 (E)$ rewrites:
\begin{eqnarray*}
{\cal L}_2(E) & = & {M\over 8} \sum_{AA'}\delta (E-E_A){1\over{E_A
-E_{A'}}}\\ & & \mbox{} {\sum_{\alpha\beta}\sum_{\alpha'\beta'}}
{{\big\vert{{C_{\alpha\beta}}^{A}}\big\vert}^2}
{{\big\vert{{C_{\alpha'\beta'}}^{A'}}\big\vert}^2}
{\delta_{\beta\beta'}}{{\big\vert<\alpha'\big\vert{p_1}
\big\vert\alpha>\big\vert}^2}
\end{eqnarray*}
In the previous expression appears the density of states 
$\rho_{\alpha\beta} (E)= \sum_A \delta (E-E_A) | C_{\alpha\beta}^A|^2 $
For the TIP-problem it was shown ~\cite{js,wp} that it is well 
described by the Breit-Wigner form:
\begin{equation}
{\rho_{\alpha\beta}}(E) = {\rho_{BW}}(E- {\epsilon_{\alpha\beta}})=
{1\over 2\pi}   {\Gamma \over
{ (E-\Gamma_0-{\epsilon_{\alpha\beta})^2+\Gamma^2/4}}}
\end{equation}
The shift $\Gamma_0$ is 
negligible for weak enough $U$ and the 
width $\Gamma$ can be estimated using the Fermi Golden rule: $ \Gamma = 2 \pi
<H_{od}^2> / \Delta_2^e$ where $ <H_{od}^2>$ is the 
variance of the non diagonal matrix elements of $H$ in the Fock basis. 
$ 1/ \Delta_2^e$ is the effective density of Fock states coupled 
by the two-body interaction. For a lattice model with $N$ sites
it is assumed that $H_{od}$ are independant
normal variables characterized by a variance of the order of $U^2/N^3$.
This estimate comes from the assumption that the one
particle wavefunctions are ergodic, as implied by the $O(N)$ invariance
in random matrix theory. This corresponds to the zero mode contribution
of a diffusion process (higher modes have been considered ~\cite{agkl}).
Within the zero-mode approximation, $\Delta_2^e
= \Delta_2 = \Delta_1^2 /B$, for $g_1 \ll 1$.
Using these approximations we obtain, 
\begin{eqnarray*}
{{\cal L}_2} (E) & = & {{\pi M} \over 4}\sum_{\alpha \beta}
\rho_{BW}(E- {\epsilon_{\alpha\beta}}) 
\sum_{\alpha'} {{\big\vert} 
<\alpha'\big\vert{p_1}\big\vert\alpha>{{\big\vert}^2} \over 
{{\epsilon_{\alpha}}- {\epsilon_{\alpha'}}}}- \\ & & \mbox{}
-{\pi \over 2}{\Gamma} \sum_{\alpha \beta}\rho_{BW}(E- 
{\epsilon_{\alpha\beta}}) 
\sum_{\alpha' \beta'}\rho_{BW}(E-
{\epsilon_{\alpha'\beta'}}) 
\\ & & \mbox{}
\delta_{\beta\beta'}
{{\big\vert}
<\alpha'\big\vert{p_1}\big\vert\alpha>{{\big\vert}^2} \over
{E - {\epsilon_{\alpha' \beta'}}}}
\end{eqnarray*}
The first term in ${\cal L}_2 (E)$ contributes to ${{C_2}^{(0)}}(E)$, i.e 
to the two-particule curvature in the abscence of interactions. The 
second part is proportional to the interactions 
so that ${\cal L} _2 (E)$ rewrites:
\begin{equation}
{\cal L} _2 (E)= {{\cal L}_2}^{(0)}(E) - {\pi \over 2} 
{\Gamma \over {{\Delta_1}^3}} I
\end{equation}
which defines both ${{\cal L}_2}^{(0)}$ and $I$. 
The sign of ${\cal L} _2 (E)$ may fluctuate from sample to sample but not 
the relative sign between the two terms of the rhs of Eq.(10).
The total two-particle curvature is:
\begin{equation}
{C_2} (E)= {{C_2}^{(0)}}(E) - {\pi \over 2}{ {B \Gamma} \over {{\Delta_1}^3}}I
\end{equation}
Using ${\Delta_2}= {{{\Delta_1}^2} \over B}$ and ${g_2}= {\Gamma \over
{\Delta_2}}$, we obtain the Eq.(2) for ${C_2} (E)$.
As discussed in the introduction, the mean value of the curvature ${C_2}$ 
is zero. We are then interested in either the typical value (when it exists) 
or in the width of the distribution. For the 
non interacting case ($\Gamma =0$), the two-particle energies are given by
$E_A = {\epsilon_{\alpha \beta}}$ 
and the typical value of the curvature ~\cite{am}
${{ {\partial^2} E_A(\phi)} \over {\partial\phi^2} } \big\vert_{\phi=0} 
\propto {g_1}{\Delta_1}$  so that ${{C_2}^{(0)}} \propto {{{g_1}{\Delta_1}}
\over {\Delta_2}}$. In the same way, replacing 
$\sum_{\alpha}$ by $\Delta_1^{-1} \int _{-B}^{B} dx $,
 the typical width of $I$ defined by its absolute value is:
\begin{equation}
{I_{typ}}  =  {{g_1}{{\Delta_1}^2} \over B} I(0,\infty).
\end{equation}
The constant $I(0,\infty)$ is given by the integral
\begin{equation} 
\int _{- \infty}^{\infty} dx \int _{- \infty}^{\infty} dy
{1 \over {(x+y{)^2} + 1}}
\int _{- \infty}^{\infty} dz {1 \over {(z+y{)^2} + 1}}
{1 \over {|z+y|}}
\end{equation}
where ${I_{typ}}$ is evaluated at the center of the band $E=0$ and in the 
limit $B/\Gamma \rightarrow \infty$ where the bandwidth is much 
larger than $\Gamma$. In this metallic limit $g_1 \gg 1$, we obtain 
for the typical two particle curvature:
\begin{equation}
{{C_2}^{(typ)}} {\simeq} {{g_1}{\Delta_1}
\over {\Delta_2}} - {g_2}{{g_1}{\Delta_1} \over B}
\end{equation}
Because of interactions, $C_2$ is reduced by a small correction. 
This agrees with numerical results ~\cite{ba} obtained 
for N spinless fermions. Since ${C_2}(E)$ can be interpreted as a measure 
of the transport along the system, we recover that repulsive 
interactions decrease the conductivity as expected when $g_1 \gg 1$. 
It could have been anticipated from the behavior of the 
TIP-energy levels $E_A(\phi)$. In the 
absence of interaction ($U =\Gamma =0$), the TIP-spectrum for 
${g_1}\gg 1$ has many level crossings. This results from the 
superposition of two independent spectra (although each of them follows 
the Wigner Dyson statistics). A finite interaction $U$
removes these level crossings and therefore reduces their curvature.
 In contrast, when ${g_1}\ll 1$, level crossings are already suppressed 
by one particle localization and a finite interaction may enhance 
the flux dependence of $E_A(\phi)$, as numerically 
observed~\cite{wmgpf}. 

 The relation (11) has been obtained by assuming a uniform 
broadening of all the Fock states. This is no longer correct when 
$g_1 \leq 1$, for scales $L_1 < L < L_2$.  There, among the $N$ Fock 
states, only a fraction $ {N_U} (\sim L{L_1}$ in $1d$) is broadened 
by the interaction. The remaining part $ {N_0} \approx N-N_U$ has a 
negligible broadening for a short range interaction since they do correspond 
to one particle states localized far away from each other. Then, 
in order to calculate ${C_2}(E)$ for $L_1 < L < L_2$, we split the sum in 
Eq.(5) into two sums corresponding respectively to $ N_0$ TIP-states close 
to a single Fock state  and $N_U$ TIP-states broadened over $g_2$ Fock states. 
The first sum corresponds to states $\big\vert {A_0}>$ which are localized 
within ${L_1} \ll L$. In the Hilbert space of these states, it is always 
possible to build a well defined position operator ~\cite{kohn} $\hat{X}$  
even for a ring geometry. Therefore, for these states the f-sum rule is 
fulfilled 
i.e. ${{ {\partial^2} E_A(\phi)} \over {\partial\phi^2} } \big\vert_{\phi=0}
=0$ and therefore
 ${C_2}^{(0)} (E)$ is zero up to terms  proportional to
$ \exp -(L/L_1)$ . Then, in 
the expression of $C_2 (E)$ remains 
only the second class of states for which the previous argument does not 
hold since they are coupled by the interaction $U$ and extended over a scale 
$L_2 > L$. These states will still contribute to $C_2(E)$ through the second
 term in the rhs of Eq. (10). Since $ g_1 \approx 0$ (f-sum rule),
the analog of Eq. (6) for one particle yields for $ M/2 {{\big\vert}
<\alpha'\big\vert{p_1}\big\vert\alpha>{{\big\vert}^2}}$ a typical value 
of order $\Delta_1$. Using this in Eq. (11)
 and making the replacements $\Delta_2 
\rightarrow \Delta_2^e \equiv B/N_U$ and 
$\Gamma \rightarrow \Gamma(\Delta_2^e)$ given by the Golden rule with a 
density $1/\Delta_2^e$, we obtain for the $N_U$ interaction-assisted 
states a typical curvature
$C_2  \simeq  {\Gamma ({{\Delta_2}^e}) \over {{\Delta_2}^e}}$ 
 when $L_1 < L < L_2$, 
 i.e. a mobility increased by the interaction. For $ L > L_2$ this 
contribution decays itself as $\exp - (L/L_2)$.  
This is nothing but the result of a decimation of the spectrum where 
all the $N_0$ TIP-states not coupled by the interaction do disappear. 
In this regime, the typical curvature $C_2 $ coincides 
with  the TIP-conductance $g_2 (L)$ defined by Imry.

% The topological curvature

 Since $C_2$ is proportional to $g_2$ only in the localized regime i.e.
 when 
the one particle mobility 
is suppressed by localization, it cannot be considered as an appropriate 
scaling parameter for the TIP-localization transition. 
For $L \ll {L_1}$ ( metallic regime), the number $g_2$ of 
consecutive TIP-levels exhibiting 
the universal Wigner-Dyson rigidity increases with $U$, but remains 
much smaller than  $C_2$ which is essentially dominated by the one 
particle kinetic part ${g_1}{{\Delta_1}\over{\Delta_2}}$. In order to get 
rid of this one particle contribution, we are led to introduce the response 
to two independent fluxes ${\phi_1}$  and ${\phi_2}$. The simplest 
generalization of a curvature which does not contain the one particle 
kinetic part is the crossed product (two-form) defined by Eq.(3).
For $U=0$, it is straightforward to show that $C =0$, since 
$\big\vert A > = \big\vert \alpha\beta >$  and ${\sum_{\alpha \beta}} 
<{{\partial \alpha}\over {\partial {\phi_1}}}\big\vert \alpha>
<\beta \big\vert {{\partial \beta}\over {\partial {\phi_2}}}>$ is real.
This is not true anymore for the interacting case. There, using Eq.(7) 
we have:
\begin{equation}
C = Im {\sum_{A}} \sum_{\alpha \beta \gamma \epsilon}
{{C_{\alpha\beta}}^{A*}}{{C_{\gamma\epsilon}}^A}
<{{\partial \alpha}\over
{\partial {\phi_1}}}\big\vert \gamma>
<\beta \big\vert {{\partial \epsilon}\over
{\partial {\phi_2}}}>.
\end{equation}
 where $C_{\alpha\beta}^A \approx -i \Gamma (({E_A}- 
 \epsilon_{\alpha\beta}) - i\Gamma )^{-1}$ (this results from the 
Lippman-Schwinger equation). The approximation previously
considered for the $C_{\alpha\beta}^A$ i.e. keeping 
only the combinations $\alpha = \gamma$ and $\beta = \epsilon$, 
for the calculation of 
$C_2$, gives a real term which does not contribute. However, by taking 
$\alpha = \epsilon$ and $\beta = \gamma$, and approximating 
$\sum_{A} \Gamma ((E_A- \epsilon_{\alpha\beta)})^2 + \Gamma ^2)^{-1}$ 
by $\Delta_2^{-1}$, we obtain:
\begin{equation}
C = {\Gamma \over {\Delta_2}} Im 
{\sum_{\alpha \beta}}
 {{{<\alpha \big\vert {{\partial H}\over
{\partial {\phi_1}}} \big\vert \beta><\beta \big\vert 
{{\partial H}\over
{\partial {\phi_2}}}}\big\vert \alpha>}\over {({\epsilon_\alpha - 
\epsilon_\beta})^2}} 
\end{equation}
where we recognize the expression of the two-form connexion 
$ V(\phi_1, \phi_2)$ (see Eq.(4))  
whose integral over a closed surface in the flux space is a geometric
 phase which is independent of the interaction.

Similar curvatures of topological origin were considered in various contexts. 
To describe the Mott transition in the one dimensionnal repulsive
 $U \geq 0$ Hubbard model, Shastry 
and Sutherland \cite{shastry} introduced two "fluxes" coupled 
respectively to the charge 
and spin degrees of freedom, for a finite density of fermions (no disorder). 
The two flux curvature measures in this case the difference between the 
charge and spin susceptibilities.
In the context of semiclassical physics, Robbins and Berry \cite{berry} did 
consider 
the semiclassical approximation of a spectral two-form similar to those 
given by Eq.(3) for situations where the classical limit corresponds 
either to integrable or chaotic systems. In the integrable case, 
the semiclassical behaviour is equivalent to the Hannay
 two-form \cite{hannay}. For non integrable systems, physical realizations 
of such a two-form is a debated question. The curvature C we introduced 
for the TIP-problem might be such an example.

This research was supported in part by the National Science Foundation 
under Grant No.PHY94-07194, by the Israel Academy of Sciences and the 
Fund for the Promotion of Research at the Technion. Useful discussions 
with Dietmar Weinmann are gratefully acknowledged.

 %          ================= REFERENCES =================


\begin{references}

\bibitem[*]{ea} Permanent address: Department of Physics, Technion, Israel 
Institute of Technology, 32000 Haifa, Israel. 

\bibitem[\#]{jlp} Permanent address: CEA, Service de Physique 
de l'\'Etat Condens\'e, Centre d'\'Etudes de Saclay, 91191 Gif sur 
Yvette Cedex, France. 
 
 
\bibitem{Shepelyansky} 
D.L. Shepelyansky, {\it Phys.Rev.Lett.} {\bf 73}, 2067 (1994).

\bibitem{Altshuler}
For a review see B.L. Altshuler and B.D. Simmons, Les Houches LXI, 1994, 
E. Akkermans, G. Montambaux, J.L. Pichard and J. Zinn-Justin editors, North 
Holland 1995.

\bibitem{agkl}
B. L. Altshuler, Y. Gefen, A. Kamenev and L. Levitov, {\it Phys. Rev. Lett.}~
{\bf 78}, 2803 (1997).

\bibitem{Imry} 
Y. Imry, {\it Europhys.Lett} {\bf 30}, 405 (1995).

\bibitem{js}
P.~Jacquod and D. L.~Shepelyansky, {\it Phys.~Rev.~Lett.}~{\bf 75}, 
3501 (1995).

\bibitem{wp}
D. Weinmann and J.L. Pichard, 
{\it Phys. Rev. Lett}~{\bf 77}, 1556 (1996).


\bibitem{fmgp}
K.~Frahm, A.~M\"uller-Groeling and J.-L.~Pichard, 
{\it Phys. Rev. Lett}~{\bf 76}, 1509 (1996). 

 
\bibitem{Thouless}
J.T. Edwards and D.J. Thouless, {\it J.Phys.} {C \bf 5}, {807} {(1972)}; 
D.J. Thouless, {\it Phys.Rep.}, {\bf 13}, 93 (1974).

\bibitem{am}
E. Akkermans and G. Montambaux,{\it Phys.Rev.Lett.} {\bf 68}, 642 (1992),
for a review see E. Akkermans,{\it Journal of Math.Phys.} {\bf 38}, 1 
(1997).

\bibitem{delande}
J. Zakrzewski and D. Delande, {\it Phys.Rev.E} {\bf 47}, 1650 (1993), 
F. von Oppen, {\it Phys.Rev.Lett.} {\bf 73},798 (1994)
and Y.V. Fyodorov and H.J. Sommers,{\it Phys.Rev.E} {\bf 51}, 2719 (1995).

\bibitem{w}
D. Weinmann, unpublished.


\bibitem{stone}
for a review see  Quantum Hall Effect, M. Stone editor, World 
scientific,(1992).


\bibitem{shapere}
for a review see Geometric phases in Physics, {\it Advanced series in 
mathematical Physics}, A. Shapere and F. Wilczek, editors, World Scientific, 
(1989).


\bibitem{ba}
R. Berkovits and Y. Avishai, {\it Europhys. Lett.} {\bf 29}, 475 (1996).

\bibitem{wmgpf}
D.~Weinmann, A.~M\"uller-Groeling, J.-L.~Pichard and K.~Frahm, 
{\it Phys. Rev. Lett.}~{\bf 75}, 1598 (1995).

\bibitem{kohn}
W. Kohn, {\it Phys. Rev.} {\bf 133}, 171, (1964) and Many Body physics,
Les Houches 1967, C.de Witt and R.Balian editors, Gordon and Breach 1968.


\bibitem{shastry}
B.S. Shastry and B. Sutherland,{\it Phys.Rev.Lett.} {\bf 65}, 243 (1990).

\bibitem{berry}
J.M. Robbins and M.V. Berry, {\it Proc.Roy.Soc.London} {A \bf 
436}, 631 (1992).

\bibitem{hannay}
J.H. Hannay, {\it J.Phys.A:Math.Gen.} {\bf 18}, 221 (1985).

\end{references}
\end{document}